\documentclass[final]{svjour2}
\usepackage{graphicx}
\usepackage{color}
\usepackage{rotating}
\usepackage{amssymb}
\usepackage{mathptmx}
\usepackage[numbers,sort]{natbib}
\makeatletter
\journalname{Journal of Low Temperature Physics}

\bibpunct{}{}{,}{n}{}{,} 

\begin{document}

\newcommand{\hdblarrow}{H\makebox[0.9ex][l]{$\downdownarrows$}-}
\title{Processing of X-ray Microcalorimeter Data with Pulse Shape Variation using Principal Component Analysis}

\author{D. Yan \and T. Cecil \and L. Gades \and C. Jacobsen \and T. Madden \and A. Miceli }

\institute{Department of Applied Physics, Northwestern University,\\ Evanston IL 60208, USA \\
Advanced Photon Source, Argonne National Laboratory, \\ Argonne IL 60439, USA \\
\email{amiceli@anl.gov}}

\maketitle

\begin{abstract}
We present a method using principal component analysis (PCA) to process x-ray pulses with severe shape variation where traditional optimal filter methods fail. We demonstrate that PCA is able to noise-filter and extract energy information from x-ray pulses despite their different shapes. We apply this method to a dataset from an x-ray thermal kinetic inductance detector which has severe pulse shape variation arising from position-dependent absorption.

\end{abstract}
  

\section{Introduction}

A common method of pulse processing for low temperature microcalorimeters is the optimal filter [\cite{Szymkowiak:1993ue}], where one cross-correlates a pulse with a pulse model (or convolves a pulse with the time reverse of a known model). This method maximizes the signal to noise under the conditions that the pulse shape and noise are stationary. However in real detectors, these conditions are not always satisfied for a variety of reasons.

We describe here the use of principal component analysis (PCA) [\cite{Malinowski:2002}] as a non-parametric analysis approach that requires no prior knowledge of the dataset for the pulse processing of low temperature detectors [\cite{Busch:LTD16}]. In this work, we draw upon a PCA-based approach used in x-ray spectromicroscopy analysis [\cite{Lerotic:2004br,Lerotic:2014ii}] to examine a simple simulated dataset consisting of pulses of different decay times and different pulse heights. We then apply our approach to a real dataset with severe pulse shape variation.  

\section{Principal Component Analysis}
\label{sec:pca_math}

When x-rays are absorbed in superconducting microcalorimeter detectors, a pulse is generated over some finite time before equilibrium is restored.
Consider a set of individually triggered detector pulses ($n = 1, \ldots N$) which are each sampled in time ($t = 1, \ldots T$), yielding a data matrix $D_{T \times N}$.  Our goal is to represent these data using a basis set $C_{T \times S}$ with $S$ characteristic pulse shape factors, with each individual pulse being represented by a weighting $R_{S \times N}$ of members of this basis set, or 
\begin{equation}
  D_{T \times N} = C_{T \times S}\cdot R_{S \times N}.
  \label{eqn:decomposition}
\end{equation}
If we can find a reduced subset with $S^{\prime}<T$ pulse shape factors that the data tell us must be present, we can represent each pulse not with all $T$ time points but in terms of its $S^{\prime}$ weighting factors.  This gives a more compact representation of a pulse over fewer variables, and once the matrix $C_{T\times S^{\prime}}$ has been determined and its matrix inverted, we can find each pulse's weighting factors $R_{S^{\prime}\times N}$ by a simple and rapidly-calculated matrix multiplication
\begin{equation}
 R_{S^{\prime} \times N} = C^{-1}_{T \times S^{\prime}} \cdot D_{T \times N}.
  \label{eqn:r_matrix}
\end{equation}
This analysis is made simpler if the matrix $C_{T \times S}$ is constructed to have orthogonal vectors (to enable matrix inversion using simple transposition) sorted in order of decreasing statistical significance (thus allowing the reduced basis set$C_{T \times S^{\prime}}$ to be easily separated from the full basis set $C_{T \times S}$).  This is precisely what is accomplished by PCA [\cite{Malinowski:2002}]. To calculate $C_{T\times S}$, we first calculate the time covariance about the origin of
\begin{equation}
Z_{T \times T} = D_{T \times N} \cdot D^T_{N \times T}	
  \label{eqn:covariance}
\end{equation}
(the relationship between PCA, SVD, and covariance matrices is discussed in textbooks on the topic [\cite{Malinowski:2002}] as well as in Appendix B of [\cite{Lerotic:2004br}]).  Because this time covariance is symmetric, we can represent it in terms of a set of eigenvectors $C_{T\times S}$ and eigenvalue weightings $\Lambda_{S\times S}$, or
\begin{equation}
	Z_{T\times T} \cdot C_{T\times S} = C_{T\times S} \cdot \Lambda_{S \times S},
\label{eqn:covariance_eigenvectors}
\end{equation} 				
where $S = T$ at the outset of our analysis. Most numerical eigenvalue-solving routines sort their output in terms of decreasing eigenvalue weightings.  As a result,
the first eigenvector (or the \emph{eigenpulse}) is essentially an average of the pulses. The second eigenvector gives the first correction to that average, the third eigenvector gives the next correction to the first two, and so on. 
Poorly correlated noise is exiled to higher order eigenvectors [\cite{CommentOnNoise}].  In this way, one can arrive at a reduced set $C_{T\times S^{\prime}}$ of eigenvectors which describe all of the significant characteristic pulse shape components, and because this is an orthonormal matrix its inverse is given by the transpose so that Eq.~\ref{eqn:r_matrix} can be calculated from the reduced set of eigenvectors as
\begin{equation}
	R_{S^{\prime} \times N} = C_{T\times S^{\prime}}^{-1}\cdot D_{T\times N} = C^{T}_{S^{\prime} \times T}\cdot D_{T\times N} 			
\label{eqn:r_from_cd}
\end{equation}
With the reduced set  of $S^{\prime}$ eigenvectors, one can also generate a compressed and noise-filtered version of the original data as
\begin{equation}
	D^{\prime}_{T\times N} = C_{T\times S^{\prime}}\cdot R_{S^{\prime}\times N}.
\label{eqn:rebuilt_data}
\end{equation}					

\begin{figure}
\begin{center}
\includegraphics[%
  width=1.0\linewidth,
  keepaspectratio]{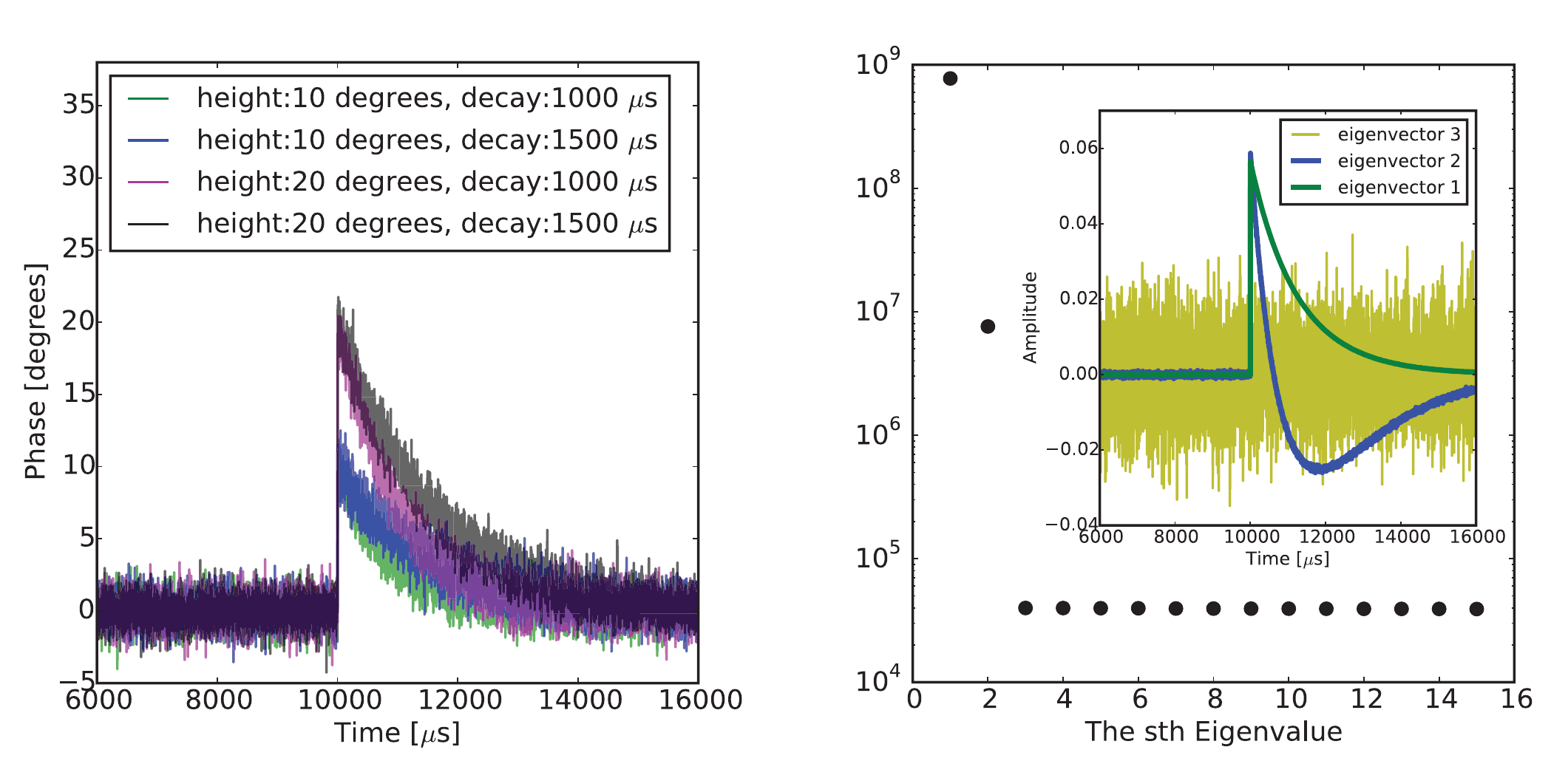}
\end{center} 
\caption{Illustration of eigenvector representation of some simulated pulses.  At {\it Left} is shown several individual pulses from the simulated dataset. They have a combination of two different heights and shapes, so there are four groups of them. The {\it Right} figure shows the eigenvalues (which are from $\Lambda_{S \times S}$), and the insert shows the first three eigenvectors. (Color figure online)}

\label{fig:fake-fig1}
\end{figure}

\begin{figure}
\begin{center}
\includegraphics[%
  width=1.\linewidth,
  keepaspectratio]{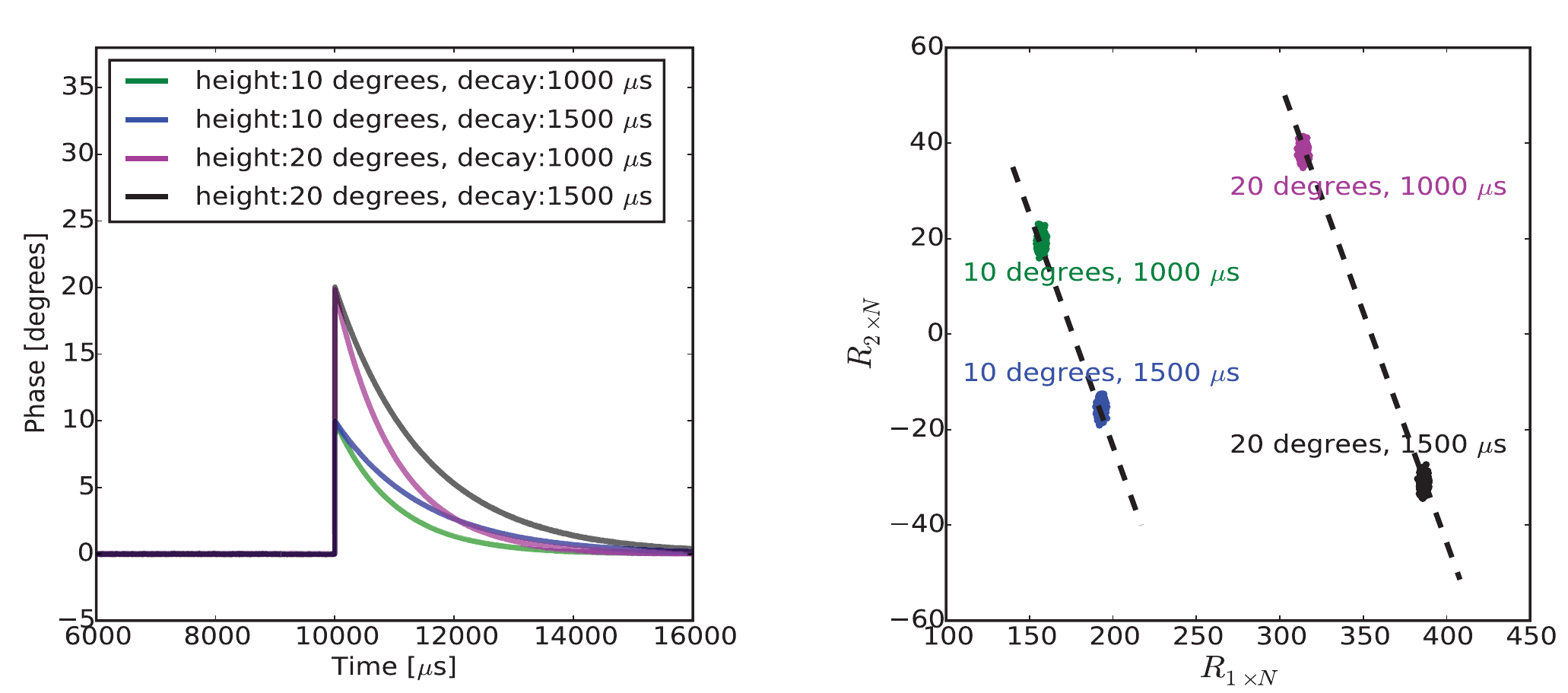}
\end{center}
\caption{The {\it Left} figure shows four PCA reconstructed pulses (Eq.~\ref{eqn:rebuilt_data}) for $S^{\prime}=2$. The {\it Right} figure represents the distribution of the elements from the weighting matrix $R_{S^{\prime} \times N}$ for $S^{\prime}=2$ from the PCA analysis of the simulated data. (Color figure online)} 

\label{fig:fake-fig2} 
\end{figure}

In order to gain intuition on how PCA treats pulse data, we have simulated a dataset which contains exponential pulses with two decay times, two pulse heights and white noise as shown in the left subfigure of Fig.~\ref{fig:fake-fig1}. As shown in the right subfigure, when decomposed this dataset contains two primary eigenvectors. The third (and higher) eigenvector contains no shape information and corresponds to noise in the dataset. Thus, we can rebuild the dataset as $D^{\prime}_{T\times N} = C_{T\times S^{\prime}=2}\cdot R_{S^{\prime}=2 \times N}$. As shown in the left subfigure in Fig.~\ref{fig:fake-fig2}, noise is greatly filtered yet the pulse shape and height features remain. The right subfigure of Fig.~\ref{fig:fake-fig2} shows the distribution of elements from the weighting matrix $R_{S^{\prime}=2 \times N}$, where components 1 and 2 respectively are the weighting factors of the 1$^{\rm st}$ and 2$^{\rm nd}$ eigenvectors.

For pulses with the same shape and height, their weighting factors are the same, so a plot of individual pulses as dots at their particular eigenvector weightings shows four clusters in the right subfigure of Fig.~\ref{fig:fake-fig2}. For pulses with same shape and different height, they have the same ratio of component 1 to component 2.  For pulses with different shape but the same height, the linear combination of their weightings is the height, so their data points are on the same line with lines that correspond to different heights parallel to each other.

\section{Analysis of TKID Data Using PCA}
\label{sec:pca_for_tkid_data}

\begin{figure}
\begin{center}
\includegraphics[%
  width=1.\linewidth,
  keepaspectratio]{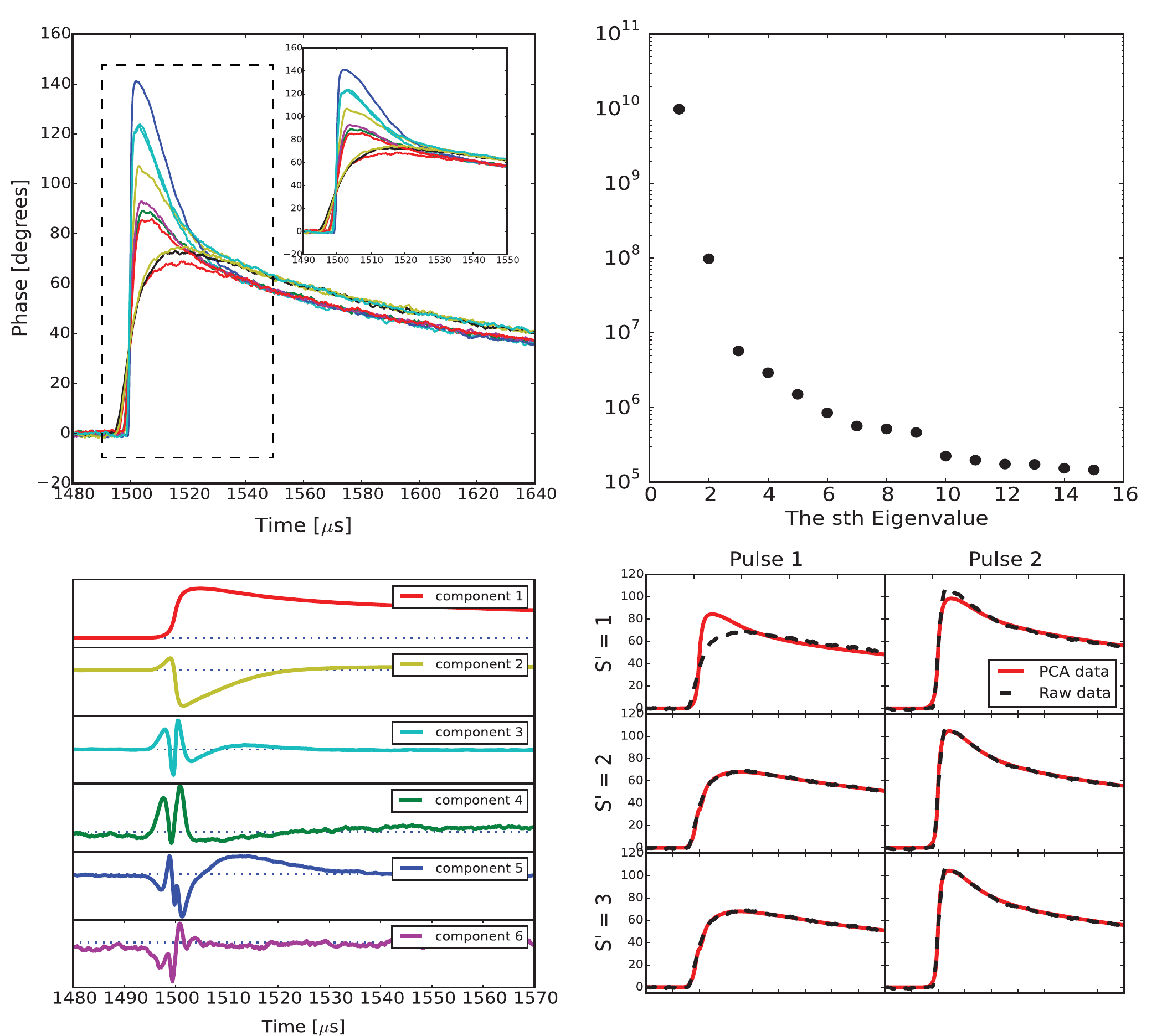}
\end{center}
\caption{Figure on the {\it Top Left} shows several individual pulses from the TKID device. A clear shape difference could be seen at the beginning. After some equilibrium time the pulses go to two branches; the lower one is Mn $K\alpha$, and the upper one is Mn $K\beta$. The Figure on the {\it Top Right} is the first fifteen eigenvalues and the {\it Bottom Left} the first six eigenvectors. The {\it Bottom Right} figure shows two raw pulses ({\it black}) in comparison with PCA reconstructed pulses (Eq.~\ref{eqn:rebuilt_data}) for $S^{\prime}=1,2,3$ ({\it red}) (Color figure online). }

\label{fig:tkid-data1}
\end{figure}

We now apply the PCA method to a real dataset from an x-ray thermal kinetic inductance detector (TKID). While other groups have reported TKIDs with 75 eV resolution at 6 keV [\cite{Ulbricht:2015kz}], we worked here with a TKID [\cite{Quaranta:2013jg, Miceli:2014kn}] from which pulse shapes were strongly dependent on the location on the sensor at which an x-ray was absorbed (see Fig.~\ref{fig:tkid-data1} {\it Top Left}). About $30 \mu$s after the start of a pulse, the pulse shape does not vary and the amplitude is proportional to the energy, the Mn $K\alpha$ and Mn $K\beta$ lines of the Fe-55 source become apparent. In such a dataset, a traditional matched or optimal filter gives no energy information, since the pulse shapes are so different that energy could not be simply extracted from pulse height or area. This has motivated us to consider a PCA analysis which makes no assumptions of the dataset. 

Following the PCA analysis presented in Sec.~\ref{sec:pca_math}, the eigenvalues and eigenvectors are calculated and shown in Fig.~\ref{fig:tkid-data1}. The first two eigenvalues are most significant, but eigenvalues 3--9 encode some subtle variations in the data. In particular, the fluctuations near a time of 1500 $\mu$s are related to the jitter in the rise time; these components are likely highly correlated with arrival time. A variant of PCA analysis (using singular variant decomposition, or SVD) recently has been studied for the detection of nearly-coincident pulses [\cite{Alpert:LTD16}]. While the components beyond the first two may show some correlation with photon energy, we restrict the analysis in this paper to the first two components for simplicity. We can see from the bottom right subfigure in Fig.~\ref{fig:tkid-data1} that there is qualitatively no large difference between $S^{\prime}=2$ and $S^{\prime}=3$, though rigorous and robust selection metrics for $S^{\prime}$ need to be developed in the future. 

\begin{figure}
\begin{center}
\includegraphics[%
  width=1.\linewidth,
  keepaspectratio]{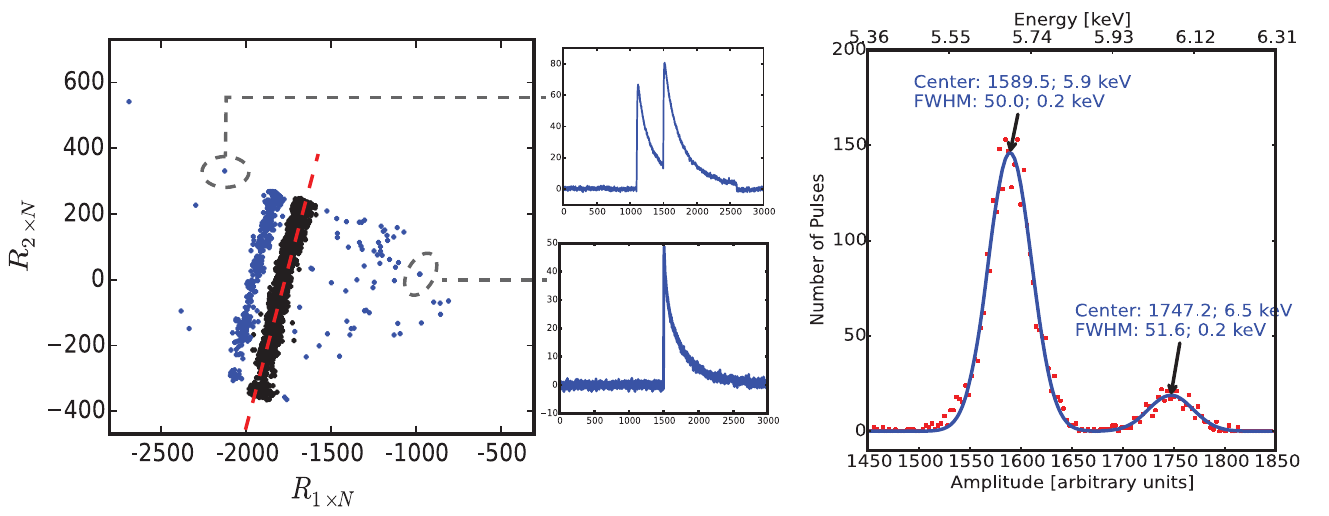}
\end{center}
\caption{The {\it Left} figure shows the distribution of elements in the weighting matrix $R_{S^{\prime}=2 \times N}$ from the PCA analysis of the TKID data. The {\it Upper} insert shows a pileup event and the {\it Lower} insert shows a low energy event, both with a position separate from the main cluster. The {\it Right} figure is the histogram of the pulse set.}

\label{fig:components_and_histogram}
\end{figure}

In order to extract energy information, we examined the weighting matrix $R_{S^{\prime}\times N}$ with $S^{\prime}=2$ which is a 2D scatter plot shown in the left subfigure of Fig~\ref{fig:components_and_histogram}.  We see two clusters which we associate with the Mn $K\alpha$ (black) and Mn $K\beta$ (blue) lines; black points are pulses in the lower $K\alpha$ branch as in the {\it Top Left} figure of Fig.~\ref{fig:tkid-data1}, and blue ones (those who are not outliers) are in the higher $K\beta$ branch. These clusters can be automatically detected and separated [\cite{Malinowski:2002}], and we have already used these automated approaches in other contexts [\cite{Lerotic:2004br,Lerotic:2014ii}].
By fitting a line (red) to the Mn $K\alpha$ cluster, we can generate an axis which was used to rotate the 2D scatter plot of the weighting matrix so that the clusters are vertical [\cite{PositionCorrelation}]. 
The projection onto the x-axis is used to generate the energy histogram in the right subfigure in Fig.~\ref{fig:components_and_histogram}. Thus, the energy can be correlated to a linear combination of the first two PCA components. 

We should note that this dataset includes pileup (i.e., more than one pulse in a single time record $T$) and low energy events. These events, shown in the insertion in the left subfigure of Fig~\ref{fig:components_and_histogram}, result in PCA weights that are vastly different, or points isolated from the main clusters.  By using $S^{\prime} > 2$  components, pileups can be further distinguished from low energy events. This suggests that PCA can be effective for pileup rejection.

\begin{figure}
\begin{center}
  \includegraphics[%
  width=1.\linewidth,
  keepaspectratio]{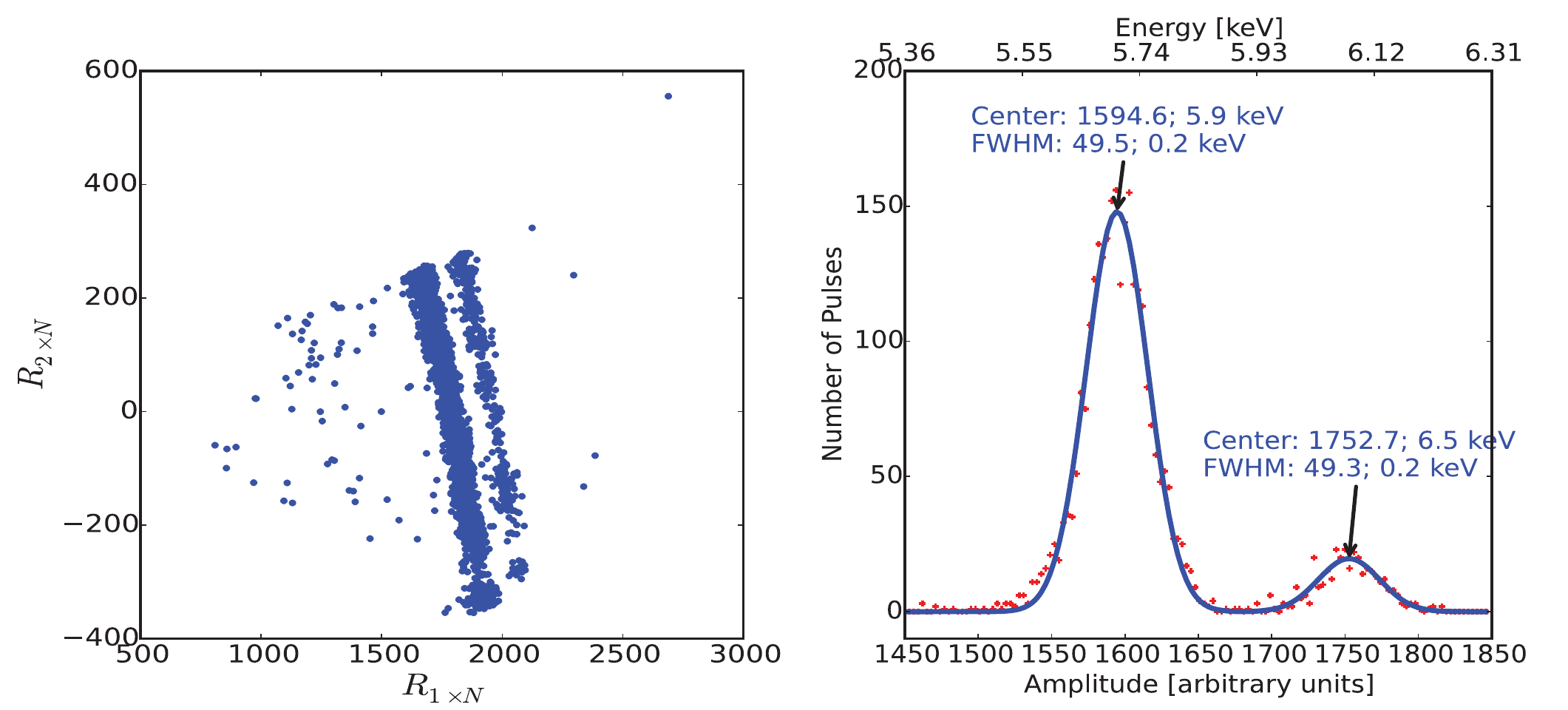}
\end{center}
\caption{The {\it Left} figure represents the entire dataset's weighting matrix data distribution, which is calculated with a training eigenvector set from 200 pulses. The {\it Right} figure is the energy histogram generated after rotating this weighting matrix.}

\label{fig:data_training}
\end{figure}

One disadvantage of PCA is its time-consuming eigenvector calculation. A solution is to use a smaller set of pulses as a training set. As an example, we used the first 200 pulses to perform the  PCA decomposition and obtain an eigenvector set $C^{training}_{T\times S^{\prime}}$. Selecting $S^{\prime} = 2$, and using Eq.~\ref{eqn:r_matrix}, we obtained the weighting matrix for the remaining $N = 3088$ pulses, which is shown in the left subfigure of Fig.~\ref{fig:data_training}. Compared to the left subfigure of Fig~\ref{fig:components_and_histogram}, despite an inverse of the first component the training data agrees well with what we obtain from direct PCA composition of the entire dataset. The energy histogram also shows very little change. With the trained set of eigenvectors, the PCA reconstruction of the data simplified to a matrix multiplication. This method could enable fast, real-time pulse processing. More work is needed to determine a sufficient number of pulses for the training set. 

\section{Conclusions}
We have introduced a non-parametric method for TKID pulse processing based on PCA, and have shown that it is beneficial for datasets with pulse shape variation. We have shown that PCA reduces data noise by the selection of a few number of components, and provides energy information by converting the data into a lower dimension basis system. Moreover, it also provides a new method to identify pileup events and for fast, real-time pulse processing. 

\begin{acknowledgements}
Use of the Center for Nanoscale Materials was supported by the U.S. Department of Energy, Office of Science, Office of Basic Energy Sciences, under Contract No. DE-AC02-06CH11357. Work at Argonne National Laboratory was supported by the U. S. Department of Energy, Office of Science, Office of Basic Energy Sciences, under Contract No. DE-AC02-06CH11357. Devices in this paper were fabricated at CNM; we gratefully acknowledge assistance from Ralu Divan, Leo Ocola, Dave Czaplewski and Suzanne Miller at CNM. We thank Mirna Lerotic and Rachel Mak for useful discussion on PCA. Finally, we thank the reviewers for their useful insights.  
\end{acknowledgements}

\bibliographystyle{JLTPv2}
\bibliography{bibtex-Nino}

\begin{thebibliography}{10}
\bibitem{CommentOnNoise}
We note that our covariance-based approach makes no reference to the spectral
  characteristics of the noise; we only ask if it is correlated with the
  signal, or not. It may be that further improvements in our analysis could be
  obtained by separating white noise from non-white noise, as suggested by an
  anonymous reviewer; we hope to explore this in future work.
\bibitem{PositionCorrelation}
Using the maximum pulse height as a proxy for photon interaction position, the
  data's new distribution along the y-axis correlates with the photon
  interaction position, and the projection onto the x-axis (i.e., energy) has a
  very weak dependence on position.
\end{thebibliography}

\end{document}